\documentclass[10pt,journal,compsoc]{IEEEtran}

\ifCLASSOPTIONcompsoc
   \usepackage[nocompress]{cite}
\else
    \usepackage{cite}
\fi

\ifCLASSINFOpdf
 
\else
 
\fi
\usepackage{multirow}
\usepackage{graphicx}
\usepackage{xcolor}
\usepackage{tabularx}
\usepackage{booktabs}
\usepackage{authblk}

\begin{document}

\title{FIDS: Fuzzy Intrusion Detection System for simultaneous detection of DoS/DDoS attacks in Cloud computing}


      \author{Saeed Ahmadi \\Master of Cybersecurity and Threat Intelligence,\\ University of Guelph\\isaeedahmadi22@gmail.com}
\date{Feb 2023}

\author{\IEEEauthorblockN{Peyman Khordadpour\IEEEauthorrefmark{1}, Saeed Ahmadi \IEEEauthorrefmark{2}}
	\IEEEauthorblockA{
	 \\
		\IEEEauthorrefmark{1}Master of Computer Engineering, Germany
 \\ peyman.khordadpor@yahoo.com	\\
 \IEEEauthorrefmark{2}Master of Cybersecurity and Threat Intelligence, \\ University of Guelph, ON, Canada
 \\isaeedahmadi22@gmail.com}
							}



\IEEEtitleabstractindextext{%
\begin{abstract}
\textcolor{black}{
In recent times, I've encountered a principle known as cloud computing, a model that simplifies user access to data and computing power on a demand basis. The main objective of cloud computing is to accommodate users' growing needs by decreasing dependence on human resources, minimizing expenses, and enhancing the speed of data access. Nevertheless, preserving security and privacy in cloud computing systems pose notable challenges. This issue arises because these systems have a distributed structure, which is susceptible to unsanctioned access - a fundamental problem. In the context of cloud computing, the provision of services on demand makes them targets for common assaults like Denial of Service (DoS) attacks, which include Economic Denial of Sustainability (EDoS) and Distributed Denial of Service (DDoS). These onslaughts can be classified into three categories: bandwidth consumption attacks, specific application attacks, and connection layer attacks. Most of the studies conducted in this arena have concentrated on a singular type of attack, with the concurrent detection of multiple DoS attacks often overlooked. This article proposes a suitable method to identify four types of assaults: HTTP, Database, TCP SYN, and DNS Flood. The aim is to present a universal algorithm that performs effectively in detecting all four attacks instead of using separate algorithms for each one. In this technique, seventeen server parameters like memory usage, CPU usage, and input/output counts are extracted and monitored for changes, identifying the failure point using the CUSUM algorithm to calculate the likelihood of each attack. Subsequently, a fuzzy neural network is employed to determine the occurrence of an attack. When compared to the Snort software, the proposed method's results show a significant improvement in the average detection rate, jumping from 57\% to 95\%.}
\end{abstract}

}

\maketitle

\IEEEdisplaynontitleabstractindextext

\IEEEpeerreviewmaketitle

\section{Introduction}
Nowadays, with the advancement of information technology and the growing needs of network users, the issue of performing computational tasks at any time and location has gained more importance. In fact, there is a need for individuals to be able to perform heavy computational tasks without the need for expensive hardware and software, using services provided by cloud computing. Cloud computing has emerged as a solution that enables users to access information and computing resources based on their specific demands \cite{b1,s2}. This model aims to reduce the need for human resources, lower costs, and increase the speed of information access, thereby being responsive to the services and needs of network users. On the other hand, cloud computing provides easy access to a flexible and configurable set of computing resources (such as networks, servers, storage spaces, and applications) based on user demand, with minimal resource management or direct server intervention, thereby addressing the needs of users and improving network scalability \cite{1,a2}.

Despite the numerous advantages of cloud computing, it has also intensified security concerns and uncertainties, which hinder the rapid adoption of cloud computing. Security and privacy preservation in cloud computing systems are crucial and challenging due to their distributed architecture and vulnerability to unauthorized access, which is one of their fundamental problems. The cloud architecture, being a distributed and open architecture, can be a suitable target for unauthorized intrusions. Therefore, cloud environments are always at risk from network attacks, with network layer attacks having the most significant impact on cloud security. In general, cloud environments, given the risks and vulnerabilities in traditional IT infrastructures, will face increased risks and vulnerabilities. It is essential for all users, both organizations and individuals, to be fully aware of the security risks in the cloud environment\cite{s3,a3,a4}. Understanding security risks and taking preventive measures helps organizations analyze the cost-effectiveness and make the use of the cloud environment mandatory for them. Since cloud computing utilizes numerous traditional and innovative technologies, it has unique and traditional security issues and concerns. Virtualization and multi-tenancy enable different users to utilize the same physical resources, leading to specific cloud security risks that need to be understood and investigated \cite{b1,a1,b2}.
Numerous strategies exist to combat cyber threats, including employing address separation protocols, spoofing internet address protocols, disrupting the domain name system, or port scanning, and offering Denial of Service (DoS) attack services. Traditional security systems like firewalls are also part of the security arsenal. However, while these methods can defend against external threats, they may lack efficiency and efficacy when faced with internal threats and sophisticated external attacks. An effective protective layer against such attacks can be provided by Intrusion Detection Systems (IDS), which are widely used to detect cyber threats. The IDS serves as an initial safeguard against possible intrusions and aims to identify and report system vulnerabilities to the system manager. IDS can be software, hardware, or a hybrid of both, automating the process of detecting unauthorized access. Data from the system or network under scrutiny is analyzed to detect any unauthorized entries, which are then reported to the network manager via email or logs. Current algorithms usually target a particular type of DoS attack, and no single model has been created to simultaneously detect all kinds of DoS attacks. Hence, the primary emphasis is on spotting irregularities in resource usage linked to DDOS and EDOS attacks. IDS alerts, however, might not always be linked to real unauthorized entries, leading to performance degradation due to existing errors. This paper aims to improve the precision of IDS in cloud computing by reducing the rates of false positives and negatives. Specifically, we aim to identify DDOS and EDOS (Economic Denial of Sustainability) attacks in cloud computing by examining traffic patterns and standard resource consumption. We hypothesize that even though attackers can create packets resembling legitimate ones during DDOS and EDOS attacks, the resource usage in such attack scenarios diverges significantly from the norm. Therefore, we investigate traffic samples and resource utilization on cloud servers to devise a suitable method to identify four types of attacks: HTTP, Database, TCP SYN, and DNS Flood. The ultimate objective of this paper is to present a universal algorithm capable of effectively handling all four attacks. We propose using a comprehensive and unique algorithm instead of employing four separate algorithms to detect these attacks. This method promises more reliable outcomes compared to algorithms that only consider metrics related to a single type of attack, as the specific metrics of one attack might be instrumental in identifying another one.

This article is organized as follows: Section 2 explores existing research in this domain. Section 3 details our proposed method, followed by an evaluation and discussion of simulation results in Section 4. Finally, Section 5 offers a conclusion and discusses potential future work.

\section{Literature Review}
Despite the numerous advantages of cloud computing, this environment has also intensified security concerns and worries, which hinder the rapid adoption of cloud computing. All users, including organizations and individuals, must be fully aware of the security risks present in the cloud environment \cite{a7,a8,b5}. In this section, we will examine the work done in this field \cite{s4,a10,a9}.

Lombardi et al. \cite{56} introduced the Advanced Cloud Protection System (ACPS), a tool for augmenting the protection of cloud resources. ACPS, which safeguards against data attacks and rogue virtual machines, ensures network and vital infrastructure security. It stands out for its transparent operations, which remain invisible to VMs. Open-source cloud systems like Eucalyptus and OpenECP were the platforms for ACPS's initial deployment. CyberGuarder, a security instrument for cloud computing deployment, was presented in \cite{57}. By extending virtual network tools, CyberGuarder provides security for network virtualization and safeguards VMs through exhaustive application integrity validation and monitoring of application-triggered system calls. In \cite{58}, Ding et al. suggested a model for defending virtual networks against falsification and spoofing attacks. The model consists of three layers: routing, firewall, and shared network. The routing layer connects the virtual and physical networks via a dedicated logical channel. The firewall layer secures the shared network against forgery attacks, while the shared network layer blocks communication between VMs from distinct virtual network channels. DCPortalsNg \cite{59} is a method that uses Software-defined Networking (SDN) for network virtualization segregation, partitioning virtual networks into different VM types to prevent DDoS attacks. This approach effectively thwarts multi-tenancy attacks in virtual networks \cite{s1}.

SnortFlow, a system for blocking cloud environment attacks, was demonstrated by Jing et al. \cite{60}. This system amalgamates features of Snort and OpenFlow systems, and its pilot implementation was trialed on Xen-based clouds. To protect cloud applications at the SaaS and PaaS layers from unauthorized access, the Diameter-AAA protocol \cite{61} was suggested. This scheme uses network-based access control to repel illicit access requests to cloud applications. Furthermore, \cite{62} saw the introduction of Security as a Service (SECaaS) in the cloud setting, a system wherein various cloud providers independently offer cloud services. SECaaS functions at all levels (SaaS, PaaS, IaaS) to ensure service security \cite{a11,a12}. Hall and Gambäck \cite{64} extended Web Service (WS) agreements to present the SecAgreement framework, which modifies standardized processes for managing Service Level Agreements (SLAs) with a security perspective. The proposed algorithm categorizes services based on risk to customers. This framework proposes a mechanism to handle SLA violations or service termination, thereby minimizing security risks post-violation/deletion. For ensuring security during runtime in addition to defending against storage attacks, an architecture was proposed in \cite{65}. It guarantees reliability, integrity, and availability for VMs during the execution phase\cite{a13,a14}. Lanjouw et al. \cite{66} introduced SBTA, a method that uses Service-oriented architecture to detect footprints and trace the origin of DDoS attacks. They also introduced a cloud filter named Cloud Filter.

In \cite{67}, a novel algorithm for detecting anomalies in a large data center based on IP structure analysis was presented. The authors used real-world traffic pattern analysis to identify IP address distribution traits in large data centers. However, this approach falls short in detecting certain DDoS attacks that exploit a specific IP address section route.
AGURI, a network monitoring tool capable of detecting DDoS attacks via long-term traffic pattern monitoring, was introduced in \cite{69}. This algorithm necessitates knowledge of the network's normal state and uses the Simple Network Management Protocol (SNMP) for this purpose.
TPANGND \cite{13} is a security framework that identifies DDoS attacks on cloud resources by 
monitoring response time anomalies. This framework leverages a Third-Party Auditor (TPA), an entity within the cloud infrastructure tasked with overseeing the compliance of Cloud Service Providers (CSPs) with Service Level Agreements (SLAs). In TPANGND, the TPA scrutinizes specific SLA metrics by sending anonymous queries to CSPs, thus keeping tabs on SLAs and identifying service response time delays\cite{a16,b3,a17}. An identification method for Low-rate cloud DDoS (LRDDoS) attacks is proposed by researchers. They employ the Fast Hartley Transform (FHT) algorithm for traffic analysis, thanks to its low computational complexity. This method can recognize LRDDoS attacks originating both internally and externally in the cloud environment, scrutinizing traffic across frequency and time domains.

Esquellani et al. suggested a mitigation strategy against EDOS attacks in cloud computing, termed EDOS-Shield. Similarly, the cost-effective EDOS Armov framework was proposed to prevent EDOS attacks on e-commerce applications in the cloud. Kodwara et al. presented a unique approach for identifying EDOS attacks, which utilizes Time Spent on a Web Page (TSP) to detect HTTP EDOS attacks. They derived this approach from the disparity between TSP during normal and attack states of a web page. However, the approach requires manual monitoring and is ineffective against other types of EDOS attacks. The system in detects HTTP DDoS attacks in cloud environments based on Information Theoretic Entropy and the Random Forest ensemble learning algorithm. The approach involves three stages: entropy estimation, data preprocessing, and classification. The algorithm begins by using a time-based sliding window to calculate the entropy of network header attributes in incoming traffic. Traffic classified as abnormal is preprocessed and normalized before being classified into normal and HTTP DDoS categories. In identifying EDOS attacks in cloud environments, an approach considers a threshold for average CPU usage and the utility function (resource utilization cost). Virtual firewalls are then utilized to mitigate this attacks\cite {b8,a18,b9}. The majority of the research conducted in the domain of DDoS and EDOS attacks has focused on devising defense mechanisms for specific attack types in cloud computing. There is comparatively less work done in identifying such attacks.

\section{Propsoed Model}

In this segment, we present a novel and distinct model designed for the identification of various DDOS and EDOS attacks, focusing on traffic recognition and resource usage. The central concept underlying this attack detection method is the notable difference in resource consumption during attack scenarios compared to normal states, even though attackers can produce packets similar to regular packets in DDOS and EDOS attacks. By taking into account metrics associated with the four types of attacks — HTTP, Database, SYN TCP, and DNS flood — this model is equipped to efficiently handle all four attack types, regardless of their occurrence timing. As a result, a unified and all-encompassing algorithm can be employed in place of different algorithms for identifying these attacks \cite{a19,a20}.

Conversely, every service in cloud computing is deployed within a distinct virtual machine. For instance, the database and web server are situated in separate virtual machines. Various attacks target different virtual machines as detailed below:

EDOS attacks, including HTTP and Database attacks: Attackers may target the virtual web server machine during an HTTP attack, while database attacks involve the virtual database machine receiving an overwhelming number of database queries per request.
DDOS attacks, like SYN TCP flood and DNS flood attacks: The SYN TCP flood attack occurs on the virtual server machine, while the DNS flood attack affects the virtual name server machine. Hence, in designing the corresponding model, we assess metrics related to traffic patterns and resource consumption in cloud server instances to formulate a suitable method for identifying all four types of attacks. The proposed methodology's block diagram is depicted in Figure 1.

   \begin{figure}
    \centering
    \includegraphics[width=8cm]{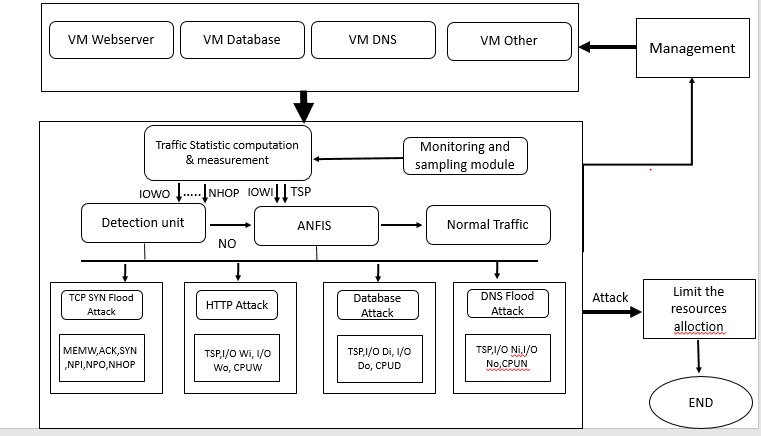}
    \caption{An overview of the proposed model}
    \label{fig:life}
\end{figure}

When virtual machines (VMs) require more resources such as memory, CPU, bandwidth, network, etc., they send requests to the hypervisor to increase their resources, and the hypervisor allocates the required resources to these VMs. If the hypervisor doesn't have sufficient resources, the VM migrates to another hypervisor. The proposed framework is placed in the middle of this process so that when a request is sent from a VM, it is first received by the measurement and traffic status calculation module. The task of this module is to extract metrics from all VMs related to the target VM. Then, the obtained results are sent to the identification unit, where the identification algorithm examines and identifies the received traffic status and detects the percentage of each type of attack.

Based on the results of this detection, if one of the four types of attacks is identified with a high percentage, the resources will be restricted. Otherwise, the resources will be allocated to the VM. The importance value of the attack percentage depends on the system's sensitivity. For more sensitive systems, even attacks with lower percentages will lead to resource limitation. In fact, the metrics used for attack identification are features extracted from traffic or resource consumption, which are introduced and described in Table 1.

\begin{table*}[ht]
\centering
\caption{Metrics for Attack Detection}
\label{tab:metrics}
\begin{tabular}{@{}ll@{}}
\toprule
Metric & Description \\
\midrule
Average request time & Measure the average time taken for a request to be processed \\
IOWi & I/O network input in the web server VM (Kbits/s) \\
IOWo & I/O network output in the web server VM (Kbits/s) \\
IODi & I/O network input in the database VM (Kbits/s) \\
IODo & I/O network output in the database VM (Kbits/s) \\
IONi & I/O network input in the name conversion server VM (Kbits/s) \\
IONo & I/O network output in the name conversion server VM (Kbits/s) \\
CPUW & CPU usage percentage in the web server VM \\
CPUD & CPU usage percentage in the database VM \\
CPUN & CPU usage percentage in the name conversion server VM \\
MemW & Memory usage percentage in the web server VM \\
R (SYN) & Relative ratio of SYN packets in TCP packets \\
R (ACK) & Relative ratio of ACK packets in TCP packets \\
R (SYN+ACK) & Relative ratio of SYN and ACK packets in TCP packets \\
NPi & Number of input packets per second \\
NPo & Number of output packets per second \\
NHOP & Number of half-open connections \\
\bottomrule
\end{tabular}
\end{table*}

Network Input in Web Server (IOWi): This measures the network I/O input in the web server virtual machine, in Kbits/s.

Network Output in Web Server (IOWo): This measures the network I/O output in the web server virtual machine, in Kbits/s.

Network Input in Database Server (IODi): This represents the network I/O input in the database virtual machine, in Kbits/s.

Network Output in Database Server (IODo): This denotes the network I/O output in the database virtual machine, in Kbits/s.

Network Input in Name Server (IONi): This shows the network I/O input in the name conversion server virtual machine, in Kbits/s.

Network Output in Name Server (IONo): This indicates the network I/O output in the name conversion server virtual machine, in Kbits/s.

Web Server CPU Utilization (CPUW): This is the percentage of CPU usage in the web server virtual machine.

Database Server CPU Utilization (CPUD): This signifies the percentage of CPU usage in the database virtual machine.

Name Server CPU Utilization (CPUN): This records the percentage of CPU usage in the name conversion server virtual machine.

Web Server Memory Utilization (MemW): This is the percentage of memory usage in the web server virtual machine.

SYN Packets Ratio (R (SYN)): This is the relative proportion of SYN packets in TCP packets.

ACK Packets Ratio (R (ACK)): This represents the relative proportion of ACK packets in TCP packets.

SYN and ACK Packets Ratio (R (SYN+ACK)): This demonstrates the relative proportion of SYN and ACK packets in TCP packets.

Packet Input Rate (NPi): This is the number of input packets per second.

Packet Output Rate (NPo): This is the number of output packets per second.

Half-Open Connections (NHOP): This is the number of half-open connections.

\subsection{Proposed Algorithm to detect various types of DDOS and EDOS attacks}
This section outlines our proposed algorithm for identifying various kinds of DDOS and EDOS attacks. The algorithm is designed to detect distinct types of assaults, such as HTTP, database, TCP SYN flood, and DNS flood, utilizing the metrics specified in Table 1. The operation of the algorithm is centered around the "status" parameter, where Stp embodies the relative total of values, t signifies the attack type (HTTP, database, TCP SYN, and DNS flood), and P denotes the P sign in each attack scenario.

To scrutinize time-based data and identify any potential changes, the CUSUM algorithm is applied [84-86-85]. Herein, we adopt a method proposed by Taylor [85], which amalgamates cumulative sums to spot shifts. The initial procedure involves accruing data points. The cumulative sum, symbolized as S0, S1, ..., Sn, is derived by evaluating the average of the data. The cumulative sum commences at zero (S0 = 0), and subsequent sums are computed by adding the discrepancy between the current value and the average to the preceding sum:

The cumulative sum computed through this method diverges from typical cumulative sum values. It takes into account the discrepancies between the values and their means. As the discrepancies between the values and their averages converge to zero, the cumulative sum invariably concludes at zero (Sn = 0).

To reinforce confidence in the changes, a self-starting analysis can be initiated to ascertain a confidence level for significant alterations. Nevertheless, an initial estimate of the change magnitude is required. In this context, Sdiff, representing the difference between Smax and Smin, is a suitable choice because it performs well for multiple changes or a robust distribution. Sdiff is calculated as follows:

For the autonomous execution of the self-starting process, the ensuing steps are followed:

First, the values of n are randomly arranged to create a self-starting sample from n units, signified as Y01, Y02, ..., Y0n.
The CUSUM self-starting run for the sample is computed and denoted as S00, S01, ..., S0n.
The maximum, minimum, and difference of the CUSUM self-starting run are evaluated as S0diff, S0min, and S0max, respectively.
The difference between the self-starting CUSUM and the original difference is compared to see if S0diff is less than Sdiff.
Please note that this translation is provided in a text format. If you require the translation in LaTeX format, please indicate so, and it will be provided as per your request.
The proposed algorithm for identifying various types of DDOS and EDOS attacks is as follows:
Define the metrics and parameters:

Stp: Relative sum of values.
t: Attack type (HTTP, database, TCP SYN, DNS flood).
P: Indicator for the presence of P sign in each attack.
Analyze the time-scheduled data using the CUSUM algorithm to detect changes [84-86-85].

Utilize a procedure introduced by Taylor \cite{85} that combines cumulative sums to identify changes.
Start by accumulating the data points. Let Yi (Y1, Y2, ..., Yn) represent the i-th data point.
Compute the cumulative sum S0, S1, ..., Sn by calculating the mean of the data.
The initial cumulative sum is S0 = 0.
Calculate the subsequent sums by adding the difference between the current value and the mean to the previous sum:
Si = Si-1 + (Yi - Ý)
where i = 1, 2, ..., n.
The calculated cumulative sum differs from regular cumulative sums as it takes into account the variances of values and means.

The cumulative sum always ends at zero (Sn = 0) since the variances of values and means become zero.
To increase confidence in detecting changes, conduct a self-starting analysis to determine a confidence level for significant changes.

Start with an estimator of the change size.
Calculate Sdiff, which represents the difference between Smax and Smin.
Sdiff = Smax - Smin = maxSi (i = 0, 1, ..., n) - minSi (i = 0, 1, ..., n).
Choose an appropriate value for Sdiff as it works well for multiple changes or a good distribution.
Execute the self-starting procedure independently:

Randomly sort the values of n to create a self-starting sample, denoted as Y01, Y02, ..., Y0n.
Calculate the CUSUM self-starting run for the sample, denoted as S00, S01, ..., S0n.
Compute the maximum, minimum, and difference of the CUSUM self-starting run as S0diff, S0min, and S0max, respectively.
Compare the difference between the self-starting CUSUM and the original difference to determine if S0diff is less than Sdiff.
By following this algorithm, different types of DDOS and EDOS attacks can be identified using the provided metrics.

   \begin{figure}
    \centering
    \includegraphics[width=5cm]{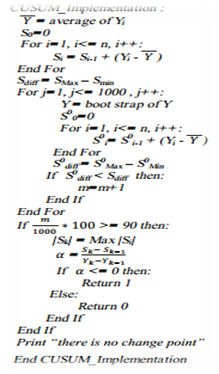}
    \caption{CUSUM algorithm}
    \label{fig:life}
\end{figure}

\subsection{HTTP Attack Signatures}
When a virtual machine (VM) requests additional resources and communicates this to the hypervisor, the hypervisor must assess the VM making the request. It should also keep an eye on the following behaviors. If the hypervisor identifies these behaviors within the VM, it signifies an HTTP attack, and the hypervisor should refrain from assigning extra resources to that VM.
The hallmarks of an HTTP attack include:

The Time Spent on Page (TSP) experiences a sudden drop during HTTP attacks.
The Network Input in the Web Server (IOWi) is significantly larger than the Network Input in the Database Server (IODi) in these attacks.
The Network Output in the Web Server (IOWo) is significantly larger than the Network Output in the Database Server (IODo) in HTTP attacks.
Web Server CPU Utilization (CPUw) is much higher than Database Server CPU Utilization (CPUD) in HTTP attacks.
The state variable for each indicator P in this attack is defined as Shttp.p. For the indicator P in an HTTP attack, Shttp.p is equal to 1 if the indicator is present, and 0 if it's absent.

\subsection{Signatures of a Database Attack}
Similar to the HTTP attack, when a VM demands more resources and relays its request to the hypervisor, the hypervisor must evaluate the VM making the requests and the VMs it's interacting with. The subsequent indicators should be scrutinized. If the hypervisor detects these behaviors in the VMs, it signifies a database attack, and the hypervisor should withhold additional resources from that VM. The distinctive signs of a database attack are typically the reverse of the signs of an HTTP attack.

Time Spent on Page (TSP) shows a sudden upsurge in database attacks.
Network Input in the Web Server (IOWi) is considerably smaller than Network Input in the Database Server (IODi) in database attacks.

\subsection{Indicators of a Database Attack}
Just like an HTTP attack, when a VM seeks more resources and communicates its request to the hypervisor, the hypervisor has to scrutinize the VM and the VMs it's interfacing with. The subsequent behaviors should be scrutinized. If the hypervisor observes these behaviors in the VMs, it implies a database attack, and the hypervisor should avoid assigning more resources to that VM. The signs of a database attack are usually inverse to those of an HTTP attack.

Time Spent on Page (TSP) shows a sudden escalation in database attacks.
Network Input in the Web Server (IOWi) is significantly smaller than Network Input in the Database Server (IODi) in database attacks.
Network Output in the Web Server (IOWo) is substantially smaller than Network Output in the Database Server (IODo) in database attacks.
The state variables for these indicators are defined as Sdatabase.p, where P denotes the indicator. For the database attack indicators, Sdatabase.p equals 1 if the indicator is detected, and 0 if it's absent.

   \begin{figure}
    \centering
    \includegraphics[width=7cm]{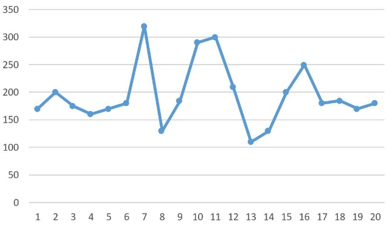}
    \caption{Reults TSP}
    \label{fig:life}
\end{figure}

\subsection{Signatures of a TCP SYN Flood Attack:}
To identify TCP SYN flood attacks, the hypervisor must pay keen attention to certain behaviors, some of which are backed by previous research while others stem from in-depth analysis and experimental studies of network traffic. These behaviors signal the potential occurrence of such cyber-attacks. Among these signs are a notable increase in the memory usage of the web server, referred to as Memw. This sharp rise is typically observed during TCP SYN flood attacks, indicating that the server's resources are being overwhelmed by a flood of requests. Another indicative behavior is a disproportionate increase in the ratio of SYN to ACK packets. Under normal circumstances, these two types of packets should roughly balance each other out. However, in the event of a TCP SYN flood attack, the number of SYN packets noticeably exceeds that of ACK packets, suggesting an attack is in progress.

An abrupt increase in the Number of Input Packets per Second (Npi) can also be a key signal of a TCP SYN flood attack. This metric represents the number of requests received by the server per second, and a sudden escalation could signify that the server is under attack.
Similarly, a sudden surge in the Number of Output Packets per Second (Npo) might hint at a TCP SYN flood attack. The Npo metric reflects the number of responses issued by the server per second, and a sudden increase could indicate that the server is trying to cope with an unusually high number of requests. Lastly, a dramatic increase in the Number of Half-Open Connections (NHOP) is a clear sign of a TCP SYN flood attack. Half-open connections are those where the SYN has been sent, but the connection is not yet fully established. An unusually high NHOP count indicates that the server is being inundated with incomplete requests, a characteristic feature of TCP SYN flood attacks. These key state variables and indicators, each representing different aspects of the server's operation, are used to monitor for possible TCP SYN flood attacks. They are derived from set procedures, providing a comprehensive means of identifying such cyber threats.

   \begin{figure}
    \centering
    \includegraphics[width=7cm]{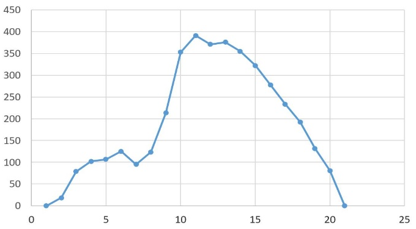}
    \caption{Reults TSP with CUSUM}
    \label{fig:life}
\end{figure}
The state parameter for indicator 1 in TCP SYN flood attack, STCP SYN Flooding 1, equals 1 if the indicator is present and 0 if the indicator is not present.
The state parameter for indicator 3, STCP SYN flooding 3, equals 1 if the indicator is present and 0 if the indicator is not present.
The state parameters for indicators 4 and 5 also equal 1 if the respective indicators are present and 0 if they are not.
\subsection{ DNS Name Resolution DDoS Attack Indicators:}

IONi is significantly higher than IODi and IOWi in DNS name resolution DDoS attacks.
IONo is significantly higher than IODo and IOWo in DNS name resolution DDoS attacks.
The CPU consumption of the VM performing name resolution is much higher than the CPU consumption of the web server and database VMs.
   \begin{figure}
    \centering
    \includegraphics[width=7cm]{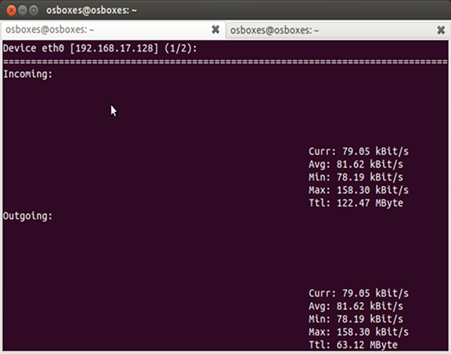}
    \caption{VM}
    \label{fig:life}
\end{figure}
\subsection{Analysis and Discusion}

The presented analysis and discussion pertains to a function whose outcomes dictate the change points in the TSP (Total Suspended Particles) values, which are critical in the detection of HTTP and database attacks. The function return values can indicate different statuses. A zero indicates K as a change point with a TSP value less than the normal state, which implies a sudden decrease. Conversely, a return value of 1 signifies a change point, with the TSP value exceeding the standard state, indicating a rapid increase. A different return value suggests that the TSP values remain stable, devoid of any significant alterations. By using indicator 1, we evaluate HTTP and database attacks, resulting in the calculations of SHTTP attack1 and Sdatabase attack1.

The CUSUM (Cumulative Sum Control Chart) algorithm, integral to this process, is depicted via sample code in Figure 2.

Following this, we apply the CUSUM algorithm to the MemW values. By doing so, we can identify potential changes and determine the change points in MemW values. Similar to TSP values, a function return of 0 indicates K as a change point with MemW values below the standard state, symbolizing a sudden decline. Conversely, a return of 1 implies K as a change point with MemW values above the regular state, signifying a sudden surge. Any other return values suggest stability in the MemW values, without significant changes. In a manner similar to the TSP values, the indicator 1 aids in recognizing TCP attacks, enabling the computation of STCP SYN flood1.
   \begin{figure}
    \centering
    \includegraphics[width=7cm]{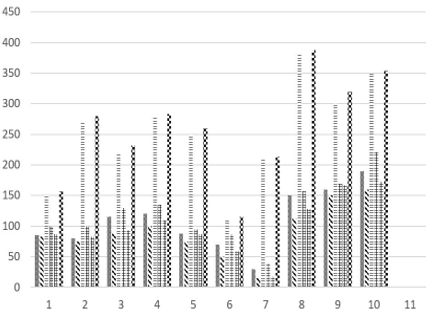}
    \caption{Traffic rate results - Output}
    \label{fig:life}
\end{figure}
   \begin{figure}
    \centering
    \includegraphics[width=7cm]{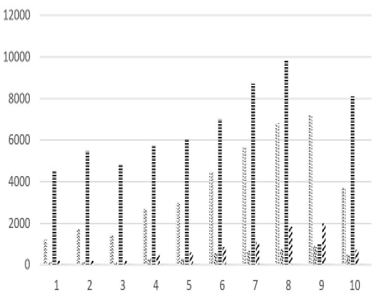}
    \caption{Traffic rate results}
    \label{fig:life}
\end{figure}
   \begin{figure}
    \centering
    \includegraphics[width=7cm]{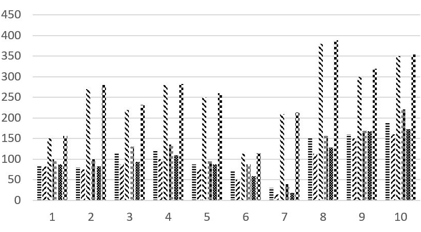}
    \caption{System overhead}
    \label{fig:life}
\end{figure}
In the next step, the NPi and NPo values are inputted separately to the execution of CUSUM to identify and calculate any changes in NPi and NPo. CUSUM is performed twice, once for NPi and once for NPo. Therefore, NPi = Y is used for the first round, and NPo = Y is used for the second round. In the first round, if the function becomes 0, it means that the NPi value at change point K is less than the normal state (sudden decrease). If the function returns 1, it means that the NPi value at change point K is greater than the normal state (sudden increase). Otherwise, there is no significant change in the NP values. Then, indicator 3 is considered for TCP attacks, and STCP SYN flood3 is calculated.

In the second round, if the function becomes 0, it means that the NPo value at change point K is less than the normal state (sudden decrease). If the function returns 1, it means that the NPo value at change point K is greater than the normal state (sudden increase). Otherwise, no significant change has occurred in the NPo values. After that, indicator 4 is considered for the attack, and STCP SYN flood4 is calculated.

Finally, the NHOP values are inputted to the execution of CUSUM to identify potential changes in NHOP. Therefore, if NHOP = 2, the function returning 0 means that the NHOP value at change point K is less than the normal state (sudden decrease). If the function returns 1, it means that K is a change point with NHOP greater than the normal state (sudden increase). Otherwise, no change has occurred. Then, indicator 5/1 in this attack is considered, and STCP SYN flood5 is calculated.
   \begin{figure}
    \centering
    \includegraphics[width=7cm]{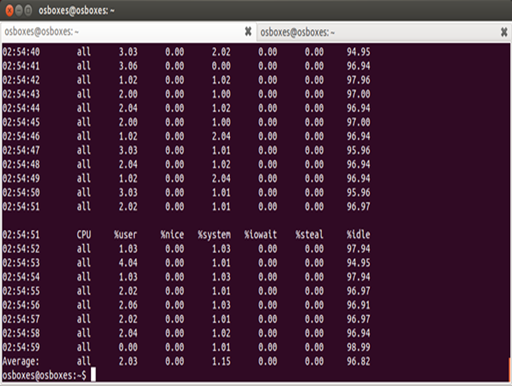}
    \caption{Config model}
    \label{fig:life}
\end{figure}
In this study, the Time since Last Web Page (TSP) metric, previously introduced by Kudora et al., is used. Any sudden changes in TSP can be detected using the CUSUM method. Figures 3 and 4 depict the TSP values in different tests and the results of executing CUSUM for TSP in HTTP and database attacks. As evident from Figure 3, identifying change points in the plot is almost impossible, but in Figure 4, identifying change points is straightforward.

   \begin{figure}
    \centering
    \includegraphics[width=7cm]{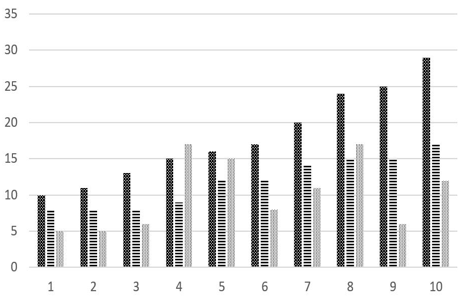}
    \caption{Attack detection results-Inputs}
    \label{fig:life}
\end{figure}
\section{ Evaluation of the proposed model}
In this section, we first discuss the extraction of metrics in the proposed model, and then the results of the detection process are examined and compared with previous results to evaluate the performance of the proposed model.
\subsection{ Simulation}
This section examines the results obtained from simulating the proposed algorithm for detecting DDOS and EDOS attacks. Here, it is assumed that the attacker is located outside the cloud, while a server of a large company is placed in a virtual machine in the cloud computing environment. VMware, a common and powerful simulator, is used for the simulation. For this purpose, Apache is used as the web server on one VM. Additionally, another VM is considered as a database server, using MySQL version 5.6.16, which is the most common and popular open-source database in the world. BIND version 9 is used for name resolution on the third virtual machine. Table 1 shows the details of the physical machine and the states of the VMs used in this study.
   \begin{figure}
    \centering
    \includegraphics[width=7cm]{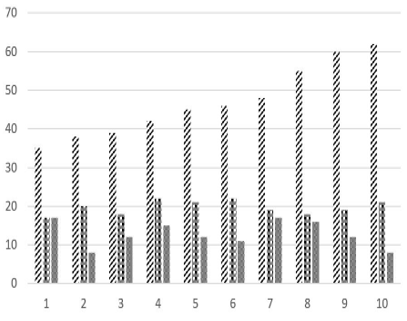}
    \caption{Memory consumption}
    \label{fig:life}
\end{figure}
In the next step, customers are considered for this company, and normal traffic is generated from customers to the server using the Curl command for sending normal traffic. In this step, commands are used to generate normal packets, and they are configured to send 20,000 packets as normal traffic for the initial load. The packet values are changed at different times during the experiment. In the next phase, the TCP SYN flood attack is performed using the hping3 tool. The hping3 command is used to create TCP SYN flood packets, and 100,000 packets are used as TCP SYN flood traffic. The number of packets is increased in subsequent attempts for a stronger attack. In the following steps, the HTTP and database attacks are executed separately using GoldenEye and JMeter tools. The GoldenEye command can be used to create an HTTP attack, and the GF key can be used to create Gt Flood attacks, while the PF key can be used to create Post Flood attacks. In the next step, the JMeter tool is used to create a higher load on the database server, assuming it as a database attack.
\subsection{Examining Network I/O in Virtual Machines of Web Server, Database Server, and Name Resolution Server}

The exploration of network Input/Output (I/O) operations has been a significant focus in previous DDOS attack studies. The primary objective of this research is to record the inbound and outbound network I/O operations in the virtual machines (VMs) hosting a web server, database server, and name resolution server. The study also aims to comprehend the pertinence of these recorded statistics. Network I/O data within the web server VM is measured utilizing Nload, an effective tool permitting detailed observation and monitoring of network activities and data. Figure 7 visualizes the acquired input and output levels of network I/O within the web server VM. Network I/O in the database server and name resolution service VMs are measured similarly using the Nload instrument
   \begin{figure}
    \centering
    \includegraphics[width=7cm]{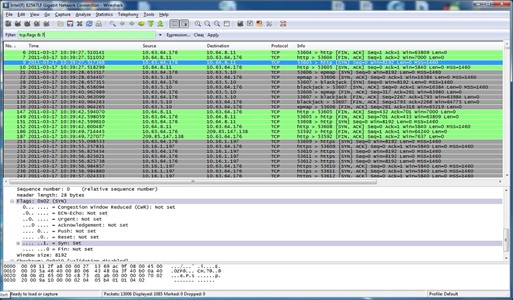}
    \caption{Traffic rate }
    \label{fig:life}
\end{figure}

Figure 8 presents the fluctuating network I/O values in the web server and database VMs under regular conditions. Figures 9, 10, 11, and 12 depict the changes during different types of attacks: HTTP, database, TCP SYN flood, and name resolution, respectively.
During an HTTP attack, the network I/O in the web server escalates sharply due to a massive influx of HTTP requests from the assailants. This surge causes the network I/O to diverge significantly from standard and idle traffic readings. In contrast, during the same HTTP attack, the network I/O in the database's VM is relatively negligible, because the primary target of the attack is the web server, leaving the database virtually untouched. This contrast is the key disparity between network I/O values during an HTTP attack and regular traffic conditions. Under normal circumstances, the network I/O in the web server VM closely mirrors that of the database VM \cite{86, 79}.

During an EDOS database attack, the network activity in the web server VM is noticeably decreased, whereas the database VM is inundated with requests. This shift results in a marked decrease in the network I/O on the web server VM, while the database VM records a comparatively high network I/O. In the event of a DNS flood attack, a deluge of name resolution requests targets the name resolution VM, rendering the web server and database VMs relatively dormant. Consequently, the network I/O values of the web server and database VMs are noticeably lower in comparison to the name resolution VM during a DNS flood attack \cite{a20,a21}. During a TCP SYN flood attack, which obstructs access to the web server, the network I/O readings for both the web server and the database VMs are minimally low, owing to the interference with regular operations.

   \begin{figure}
    \centering
    \includegraphics[width=7cm]{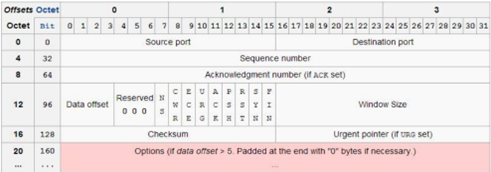}
    \caption{Packet header}
    \label{fig:life}
\end{figure}

   \begin{figure}
    \centering
    \includegraphics[width=7cm]{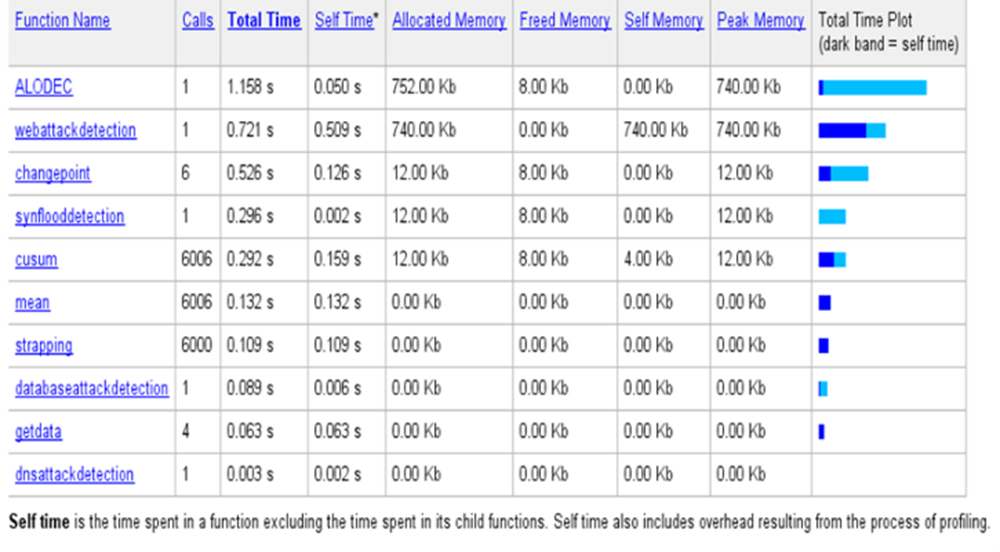}
    \caption{ANFIS results }
    \label{fig:life}
\end{figure}

\subsection{CPU Utilization in the Web Server, Database Server, and Name Resolution Virtual Machines}

Previous studies \cite{69,67,58,56} have pointed out that CPU consumption is another resource impacted during DDOS and EDOS attacks. The method proposed by Shai et al. \cite{2,1} has been adopted to gather CPU usage data. The Sar command in Linux is employed to obtain CPU usage statistics. Figure 13 visualizes the results obtained using the Sar command.
These figures represent varying CPU usage values across the web server, database server, and name resolution virtual machines during normal operating conditions as well as during HTTP, database, TCP SYN flood, and name resolution attacks.

Typically, the CPU utilization in the web server and database virtual machines mirrors the network I/O values in the database virtual machines during normal conditions and web server and database attacks. CPU consumption in the web server virtual machine closely aligns with that in the database virtual machine. CPU consumption in the web server exhibits a slight increase during an HTTP attack since the server is inundated with HTTP requests requiring substantial processing. Conversely, during an HTTP attack, the database virtual machine registers negligible CPU usage as it is not the primary target of the assault, and hence, remains relatively idle. Database attacks present a contrasting trend compared to HTTP attacks. During EDOS database attacks, the database virtual machine experiences a high volume of requests, while the web server virtual machine is underutilized. Consequently, the CPU usage percentage is lower in the web server VM but high in the database VM during the attack. During a DNS flood attack, both the web and database servers display a high CPU usage ratio, and the name resolution server also demonstrates increased CPU usage. Additionally, during a TCP SYN flood attack.

Another metric, memory usage in the web server, can be tracked using the "TOP" command in the hypervisor. This metric is relevant as prior research has established that DDOS attacks can influence memory consumption. To evaluate this value under normal conditions and to detect abrupt shifts in memory usage, both attacks and normal traffic are simulated in the system. Figure 19 depicts the memory usage values (MemW) under different test scenarios. Additionally, the results of applying the CUSUM method to MemW during a TCP SYN flood attack are presented in Figure. During HTTP, database, and DNS attacks, memory usage patterns are almost identical to those under normal conditions. However, during a TCP SYN flood attack, memory consumption can rise sharply as this attack directly influences memory use. Consequently, a sudden spike in memory usage is witnessed during a TCP SYN flood attack. While detecting change points is challenging as illustrated in Figure 19, the task becomes straightforward as shown in Figure 5-18. Figure 20 shows a sudden change between test numbers 12 and 13. Furthermore, since the slope of the CUSUM chart before this point (test number 12) is descending, it signals a change point with a value higher than the normal state for the TCP SYN flood attack (a sudden increase). No significant change points are noted in memory usage for HTTP, database, and name resolution flood attacks compared to normal conditions.

\subsection{Packet Data Analysis}
Understanding the dynamics of network packet data can be instrumental in identifying potential cyber threats such as attacks. This analysis becomes particularly crucial as the quantity of network packets may exhibit dramatic fluctuations depending on the intensity and nature of the attack, or even during periods of high-intensity standard traffic. Under typical traffic conditions, the ratio between various types of packets remains relatively consistent regardless of the intensity of traffic. However, this balance can be substantially disrupted in the event of an attack. For example, during a Distributed Denial of Service (DDoS) attack that specifically employs a TCP flow assault, the ratios of TCP ACK, TCP SYN, and TCP (SYN+ACK) packets would deviate from the norm.

To provide a detailed analysis, certain key metrics are considered:

R(SYN): The proportion of SYN packets in the total TCP packets.
R(ACK): The proportion of ACK packets in the total TCP packets.
R(SYN+ACK): The proportion of (SYN+ACK) packets in the total TCP packets.
These metrics can be effectively evaluated using a tool like Wireshark, which enables sampling packets and extracting traffic data. Figure 21, for example, demonstrates how useful data from different packet flags can be drawn from a traffic sample and analyzed. Certain critical values, such as R(ACK), R(SYN), and R(FIN), are obtained through this scrutiny.

In addition to the ratios, the packet count per second, both inbound and outbound, is also a valuable metric. The Nload tool can be effectively used to collect this data. Research [80] points out that packet count per second can exhibit sudden changes during EDoS and DDoS attacks. The packet count values in various test scenarios are illustrated in Figures 23 and 24, while Figures 25 and 26 present the application of CUSUM for pinpointing changes during a TCP SYN flood attack. Analysis of these metrics in different scenarios suggests interesting insights. For instance, an HTTP attack is associated with a sudden increase in packet count, but this value won't be significantly different during busy, normal traffic. On the other hand, a database attack sees a decrease in packet count, but the deviation from less busy, standard traffic is still marginal. The most substantial difference in packet count appears during a TCP SYN flood attack when compared to standard traffic, with a significant increase in count noted \cite{a23,a22}.

\subsubsection{Assessment of Half-Open Connections}
The tally of half-open connections serves as a vital metric used exclusively in pinpointing TCP SYN flood attacks, employable via the "netstat" command in Linux. As depicted in Figure 27, the quantity of half-open connections during an HTTP attack and a database attack aligns with the standard scenario. However, during a TCP SYN flood assault, this parameter exhibits a sudden increase.

In a TCP SYN flood assault scenario, assailants transmit an overwhelming volume of SYN packets to the targeted party, failing to reply to the ACK packets dispatched by the target server. Consequently, the three-way handshake remains incomplete, and the connection is prematurely terminated without being established, leading to a surge in half-open connections where the SYN+ACK packets are lost. This sudden shift during a TCP SYN flood attack at test 12 is visually represented in Figure 29.

\subsection{Evaluation of Precision}
The precision and reliability of the proposed algorithm are assessed by testing our identification model and juxtaposing it with preceding works. For this aim, we chose Snort, an open-source Intrusion Detection System (IDS) capable of real-time traffic scrutiny and attack identification \cite{a22,a23}. The outcomes highlight that our model executes with a superior degree of accurate identification compared to older IDS systems like Snort. A comparison of the accuracy of our model and Snort IDS is visually represented in Figure 30.

As the figure illustrates, the average precision of our model hovers around 72\%, while the average accuracy of Snort is roughly 58\%. Furthermore, the memory utilization for implementing the CUSUM algorithm has been computed, with the results exhibited in Figure 11.
The execution time for the suggested algorithm is estimated at 0.190851 seconds, and the CPU utilization is roughly 2\%, as presented in Figure 12. Leveraging the obtained information and results, the false positive rate and true positive rate have been deduced. Figure 13 showcases the graphs of training data and inference data. The left side depicts the input training data fed into the system. As per this graph, traffic is denoted as zero for standard traffic, one for web attacks, two for database assaults, three for SYN flood attacks, and four for DNS flood attacks. The right side displays the output inference values based on the training data.
Figure 14 visualizes the testing data and inference data graphs. The input testing data to the system is displayed on the left side. Similar to the preceding graph, traffic is denoted using the same values. The output inference values, based on the testing data, are showcased on the right side. The discrepancy between the inference output and the actual output is shown.

\section{Conclusion}

Safety remains a paramount concern in the realm of cloud computing. In this paper, we scrutinized Distributed Denial of Service (DDOS) attacks, with a particular focus on a variant known as TCP SYN flood attacks. These attack forms are prevalent in cloud computing. Economic Denial of Sustainability (EDOS), a unique and emergent strain of DDOS attacks, can also compromise the cloud via a DDOS onslaught, leading to cost ramifications for both consumers and cloud service providers. This study appraised HTTP and database attacks as specific examples of EDOS and DDOS attacks. The foremost step towards mitigating an attack and shielding the system against it lies in crafting a robust and resilient framework for attack detection. Numerous algorithms exist for pinpointing DDOS attacks in traditional (non-virtualized) systems. However, these algorithms fall short in a cloud setting unless appropriate modifications are implemented. Conversely, most current DDOS detection algorithms in cloud computing are reliant on packet information. Given that the packets in a DDOS attack resemble those under normal conditions, these algorithms don't offer the necessary distinctiveness and validity for identifying and detecting DDOS attacks. Furthermore, they become ineffective in the cloud milieu if requisite alterations are not made. Current solutions for EDOS attacks are generally specific to a unique attack type, and thus far, no model has been suggested for the detection of all attack varieties. In this research, we propose a comprehensive and novel algorithm for identifying various types of DDOS and EDOS attacks. This model shows effectiveness against all four attack types: HTTP, database, TCP SYN flood, and DNS flood, irrespective of their occurrence timing. Hence, a single generalized and comprehensive algorithm can be deployed in place of four separate detection algorithms. We performed simulation results using metrics and graphs, and each metric was evaluated using the CUSUM algorithm. Lastly, we compared our research findings with Snort, wherein the average detection rate with Snort hovers around 57\%, while the average detection rate via our proposed model is approximately 72\%. The data amassed was further used for traffic and existing attack repetition analysis, with the information and metrics serving as input to the inference. We then scrutinized the output information and verified its accuracy. On average, we observed an improvement in inference by 34.97

\bibliographystyle{IEEEtran}
\bibliography{Ref}

\end{document}